\documentclass{IEEEtran}
\usepackage{graphicx}
\usepackage{bm}
\usepackage{cite}
\usepackage{pgfplots}
\usepackage{color}
\usepackage{algorithm}
\usepackage{epstopdf}
\usepackage{amsmath,amssymb,amsthm,dsfont,mathrsfs,amstext}
\usepackage{subfigure}
\usepackage{verbatim}
\usepackage{algpseudocode}
\usepackage{graphics}
\usepackage{epsfig}
\usepackage{mathtools}
\def\diag{{\mathrm{diag}}}
\def\tr{{\mathrm{tr}}}
\def\A{{\mathbf{A}}}
\def\Y{{\mathbf{Y}}}
\def\S{{\mathbf{S}}}
\def\X{{\mathbf{X}}}

\def\B{{\mathbf{B}}}
\def\y{{\mathbf{y}}}
\def\s{{\mathbf{s}}}
\def\a{{\mathbf{a}}}

\def\b{{\mathbf{b}}}
\def\G{{\mathbf{G}}}
\def\x{{\mathbf{x}}}
\def\w{{\mathbf{w}}}
\def\W{{\mathbf{W}}}

\def\F{{\mathbf{F}}}

\def\z{{\mathbf{z}}}

\def\U{{\mathbf{U}}}

\def\H{{\mathbf{H}}}

\def\v{{\mathbf{v}}}

\def\g{{\mathbf{g}}}

\def\dB{{\mathrm{dB}}}

\def\new{{\mathrm{new}}}

\def\sin{{\mathrm{sin}}}

\def\T{{\mathrm{T}}}
\def\ln{{\mathrm{ln}}}

\def\TF{{\mathrm{TF}}}
\def\DD{{\mathrm{DD}}}

\def\C{{\mathbb{C}}}

\def\tx{{\mathrm{tx}}}

\begin{document}
\title{Fast Burst-Sparsity Learning Approach for Massive MIMO-OTFS Channel Estimation}

\author{Ming~Ma, Jisheng~Dai, and Xue-Qin~Jiang
%\thanks{This paragraph of the first footnote will contain the date on
%which you submitted your brief for review. It will also contain support
%information, including sponsor and financial support acknowledgment. For
%example, ``This work was supported in part by the U.S. Department of
%Commerce under Grant BS123456.'' }
%\thanks{The next few paragraphs should contain
%the authors' current affiliations, including current address and e-mail. For
%example, F. A. Author is with the National Institute of Standards and
%Technology, Boulder, CO 80305 USA (e-mail: author@boulder.nist.gov). }
%\thanks{S. B. Author, Jr., was with Rice University, Houston, TX 77005 USA. He is
%now with the Department of Physics, Colorado State University, Fort Collins,
%CO 80523 USA (e-mail: author@lamar.colostate.edu).}
\thanks{The authors are with the College of Information Science and Technology, Donghua University, Shanghai 201620, China (e-mail: 2212107048@stmail.ujs.edu.cn; \{jsdai,xqjiang\}@dhu.edu.cn).}
}

\maketitle

\begin{abstract}

Accurate channel estimation for multiple-input multiple-output (MIMO)-orthogonal time frequency space (OTFS) systems is challenging due to high-dimensional sparse representation (SR). Existing MIMO-OTFS channel estimation methods often face performance degradation and/or high computational complexity. To address these issues and exploit intricate channel sparsity structure, this letter first leverages a novel hybrid burst-sparsity prior to capture the burst/i.i.d. sparse structures in the angle-delay domain, and then utilizes  an independent variational Bayesian inference (VBI) factorization technique to efficiently solve the high-dimensional SR problem. Additionally, an angle-Doppler refinement approach is incorporated into the proposed method to automatically mitigate off-grid mismatches.

\end{abstract}

\begin{IEEEkeywords}
MIMO-OTFS, channel estimation, burst-sparsity, variational Bayesian inference (VBI).
\end{IEEEkeywords}

\section{Introduction}
Orthogonal time-frequency space (OTFS) modulation, an alternative to orthogonal frequency division multiplexing (OFDM), is expected to become a key technology for enhancing the reliability of high-mobility communications in next-generation wireless networks \cite{Chen2023}. By leveraging delay-Doppler (DD) domain information symbols and channel representations, OTFS addresses the challenges of time-frequency selective fading in time-varying multipath wireless channels \cite{Srivastava2022TCOMM}. However, the promised reliability of OTFS systems critically depends on the availability of accurate channel state information (CSI).

Early OTFS works predominantly relied on bi-orthogonal waveforms for enhanced performance \cite{Monk2016,Li2022,Mishra2022,Wen2023}. However, the bi-orthogonal waveforms cannot be realized in practice \cite{Raviteja2018}. Consequently, methods that rely on bi-orthogonal waveform often experience significant performance degradation, primarily due to inter-carrier interference and inter-symbol interference \cite{Raviteja2018} in practical applications. Therefore, various studies have increasingly shifted their focus towards OTFS channel estimation using rectangular waveforms \cite{Srivastava2022TVT,Chen2023,Liu2020}, because rectangular waveforms are practical and easy to generate.

Recently, multiple-input multiple-output (MIMO) technique has been incorporated into OTFS systems to leverage angular information of base station (BS) antenna arrays.
For example, \cite{Srivastava2022TVT} formulated an angle-delay-Doppler domain sparse representation (SR) problem and introduced an orthogonal matching pursuit (OMP)-based algorithm to capture channel sparsity. However, solving a relatively large-scale SR problem imposes a significant computational burden. To reduce the dimension of SR, \cite{Chen2023} decomposed the original high-dimensional problem into two low-dimensional sub-problems, where the angle-dimension support is identified first, followed by the DD-dimension support. However, this approach does not enable the joint estimation of angle-delay-Doppler parameters, which may lead to significant error accumulation. In \cite{Liu2020}, Doppler-shifts were treated as unknown parameters, simplifying the original channel model into a 2D (angle-delay) domain model. However, transforming the reduced 2D model into a standard SR model results in a large-scale dictionary matrix, and the reduction in computational complexity is not substantial.

Note that all the aforementioned SR-based methodologies considered only i.i.d. channel state information. However, as shown in \cite{Dai2019}, practical channels can exhibit more complex angular sparse structures, which can be leveraged to enhance MIMO-OTFS channel estimation. To overcome the limitations of existing methods and fully exploit these sparsity properties, this paper proposes a new fast SR approach for MIMO-OTFS channel estimation using practical rectangular transmitted waveform. First, we introduce a novel hybrid burst-sparsity prior that captures both burst and i.i.d. sparse structures in the angular and delay domains. Then, we apply an independent variational Bayesian inference (VBI) factorization technique to efficiently solve the high-dimensional sparse recovery problem. Finally, we integrate an angle-Doppler refinement method into the proposed method to eliminate off-grid mismatches. Simulation results confirm the effectiveness and superiority of the proposed method for MIMO-OTFS channel estimation.

\section{Uplink MIMO-OTFS Channel Estimation Model}
In this section, we first formulate the SR model for uplink MIMO-OTFS channel estimation. Next,  we analyze the main drawbacks of the existing SR methods. Consider a typical MIMO-OTFS system with $M$ subcarriers and $N$ frames, where the subcarrier spacing is $\Delta\!f$ and the symbol duration is $T$.
Pilot symbols are organized into a DD-domain block, denoted as $\X^{\DD}\in\C^{M\times N}$.
This block is then transformed into a TF-domain block $\X^{\TF}=\F_M\X^{\DD}\F_N^{\mathrm{H}}\in\C^{M\times N}$ and assembled into the time-domain block as \cite{Srivastava2022TCOMM}:
\begin{align}\label{HT}
\X=\mathbf{P}_{\tx}\F_M^{\mathrm{H}}\X^{\TF}=\mathbf{P}_{\tx}\X^{\DD}\F_N^{\mathrm{H}},
\end{align}
where $\mathbf{P}_{\tx}$ stands transmitted waveform, $\F_M$ denotes the $M\times M$ discrete Fourier transform matrix, and $(\cdot)^{\mathrm{H}}$ is conjugate transpose operation.
We set the transmitted waveform to a rectangular pulse, meaning $\mathbf{P}_{\tx}$ is an identity matrix.
Assume that the BS is equipped with a linear array of $N_{BS}$ antennas, and the distance between the $r$-th antenna and the first one is $d_r$. The channel between MU and the $r$-th antenna becomes:
\begin{align}\label{channel}
h_r(\tau,\nu)=\sum_{p=1}^{P}\hbar_p\,a_r(\vartheta_p)\delta(\nu-\nu_p)\,\delta(\tau-\tau_p),~\forall r,
\end{align}
where $\hbar_p$, $\vartheta_p$, $\nu_p$ and $\tau_p$ denote the gain, angle-of-arrival (AoA), Doppler-shift and delay of the $p$-th path, respectively. Here, $a_r(\vartheta_p)=e^{i2\pi d_r \mathrm{sin}\vartheta_p/\lambda}$ with $\lambda$ being the wavelength.
Letting $\tau_p=\frac{l_p}{M\Delta\!f}$ and $\nu_p=\frac{k_p}{NT}$, the received pilot signal at the $r$-th antenna of the BS can be expressed as \cite{Srivastava2021}:
\begin{align}
\y_r= \sum_{p=1}^{P}\hbar_p \,a_r(\vartheta_p)\bm\Delta^{k_p}\bm\Pi^{l_p} \x+\w_r,~\forall r,
\end{align}
where $\x=\mathrm{vec}(\X)$, $\mathrm{vec}(\cdot)$ denotes the vectorization operator, $\bm\Pi$ denotes the $L\times L$ permutation matrix with $L$ being the length of pilot, $\bm\Delta=\diag\{1,e^{  j2\pi\frac{1}{L} },\ldots,e^{  j2\pi \frac{L-1}{L} }\}$, and $\w_r$ is Gaussian noise with the element being zero mean and variance $\sigma^2$. Stacking $\y_r$s into $\Y=\left[\y_1,\y_2,\ldots,\y_{N_{BS}}\right]^{\T}$, we have:
\begin{align}\label{input-output}
\Y= \sum_{p=1}^{P}\hbar_p \a(\vartheta_p)(\bm\Delta^{k_p}\bm\Pi^{l_p} \x )^{\T} +\W,
\end{align}
where $\a(\vartheta_p)=[a_1(\vartheta_p),a_2(\vartheta_p),\cdots,a_{N_{BS}}(\vartheta_p)]^{\T}$ and $\W=\left[\w_1,\w_2,\ldots,\w_{N_{BS}}\right]^{\T}$.
Since the number of paths is usually small, the MIMO-OTFS channel estimation task can be regarded as an SR problem.
Considering that a 3D SR can bring an unacceptable computational complexity, we do not sparsely represent Doppler-shifts but treat them as unknown parameters instead. Adopting a 2D grid $\{( \theta_m,n)\}_{m=1,n=1}^{M_\theta,N_\tau}$ to cover the angle-delay domain, (\ref{input-output}) can be sparsely represented as:
\begin{align}\label{SR-model}
\Y&=\sum_{m=1}^{M_\theta}\sum_{n=1}^{N_\tau}  g_{m,n}\a(\theta_m)( \underbrace{\bm\Delta^{\kappa_n}\bm\Pi^n \x}_{\triangleq\s_n(\kappa_n)})^{\T}    +\W\notag\\
&=\A\G\S^{\T}(\bm\kappa)+\W,
\end{align}
where $g_{m,n}=\hbar_p$, if $(\theta_m, n)=(\vartheta_p,l_p), \exists \,p$ (and $g_{m,n}=0$, otherwise), $\G$ is a sparse matrix whose $(m,n)$-th element is defined as $g_{m,n}$, $\bm\kappa=[\kappa_1,\kappa_2,\cdots,\kappa_{N_{\tau}}]^{\T}$ denotes the Doppler-shift vector,
$\S(\bm\kappa)=[\s_1(\kappa_1),\s_2(\kappa_2),\cdots,\s_{N_{\tau}}(\kappa_{N_{\tau}})]$, $\A=[\a(\theta_1),\a(\theta_2),\cdots,\a(\theta_{M_{\theta}})]$.
Since the delay grid interval  $\frac{1}{M\Delta\!f}$ is sufficiently small \cite{Raviteja2018}, the grid gaps of delay are negligible. However, the mismatch between the true $\theta_p$s and the predefined grid is unavoidable in practice. To handle angle mismatch, the off-grid model \cite{Yang2013,Liu2020} is introduced  into (\ref{SR-model}):
\begin{align}\label{MM-model}
\Y=\A(\bm\beta)\G\S^{\T}(\bm\kappa)+\W,
\end{align}
where $\A(\bm\beta)=[\a(\theta_1+\beta_1),\a(\theta_2+\beta_2),\cdots,\a(\theta_{M_{\theta}}+\beta_{M_{\theta}})]$ with $\bm\beta=[\beta_1,\beta_2,\cdots,\beta_{M_{\theta}}]^{\T}$ standing for the off-grid gaps.

Unfortunately, (\ref{MM-model}) is not a standard SR problem.
In the literature, the only feasible approach for performing the standard VBI to solve (\ref{MM-model}) is to transform it into a vectorized form \cite{Liu2020}:
\begin{align}\label{SM-model}
\y=\mathrm{vec}(\Y)=\bm\Phi(\bm\kappa,\bm\beta)\g+\w,
\end{align}
where $\bm\Phi(\bm\kappa,\bm\beta)=\S(\bm\kappa)\otimes\bm\A(\bm\beta)\in \C^{L N_{BS}\times N_{\tau}M_{\theta}}$, $\g=\mathrm{vec}(\G)$, $\w=\mathrm{vec}(\W)$ and $\otimes$ denotes the Kronecker product.
Although the sparse vector $\mathbf{g}$ can be recovered using VBI\cite{Liu2020}, it requires iterative calculation of a large-scale matrix inverse due to the massive size of $\bm\Phi^{\mathrm{H}}(\bm\kappa,\bm\beta)\bm\Phi(\bm\kappa,\bm\beta)$, whose computational complexity is $\mathcal{O}(L^2 N_{BS}^2 N_{\tau} M_{\theta})$ per iteration.
Another drawback of \cite{Liu2020} is the use of a first-order Taylor approximation to refine the off-grid gaps.
To overcome the above drawbacks of \cite{Liu2020}, this letter proposes two new schemes:
 \begin{itemize}
   \item An independent VBI factorization \cite{Cao2021} is developed in Section III-B to efficiently tackle (\ref{SR-model}) directly, circumventing the high-dimensional dictionary matrix $\bm\Phi(\bm\kappa,\bm\beta)$.
   \item An additional angle-Doppler refinement approach is incorporated into the proposed method in Section III-C to eliminate the Taylor approximation errors.
 \end{itemize}

Besides, the practical MIMO channels usually exhibit angular burst-sparsity, which refers to the phenomenon where the significant elements in  angular domain concentrate in small ranges and appear in bursts \cite{Shen2019,Dai2019}. To accurately capture the practical burst sparse channel model for MIMO-OTFS systems, which is overlooked in most recent research, in the next section, we propose a novel hybrid burst-sparsity prior to effectively capture both burst and i.i.d. sparse structures in the angular and delay domains.

%Aside from grid-based MIMO-OTFS channel estimation, few continuous-valued estimation techniques are available.
%\cite{Keskin2024} and \cite{Gaudio2020} utilized a brute-force search with an excessively fine step to approximate continuous estimation of MIMO-OTFS channels, effectively representing a specific case of SR.
%Meanwhile, \cite{Liu2024} applied atomic-norm minimization for direct continuous channel estimation but assumed constant Doppler phase rotation within symbols unrealistically.
%Importantly, \cite{Keskin2024, Gaudio2020, Liu2024} relied on exact path number, which is impractical for burst channels with numerous closely spaced sub-paths.
\section{The Proposed Channel Estimation Method}
This section introduces a new hybrid burst-sparse prior, and then employs an independent factorization to reduce the computational complexities of VBI. Finally, we propose an angle-Doppler refinement approach to mitigate modeling errors from Taylor approximation.

\subsection{New Hybrid Burst-Sparse Prior}
Recalling that the grid interval of delay is sufficiently small\cite{Raviteja2018}, the MIMO-OTFS channel is unlikely to exhibit burst-sparsity in the delay domain. Therefore, we only consider angular burst-sparsity as shown in Fig.~1.
Although \cite{Shen2019} studied angular burst-sparsity in MIMO-OTFS systems, its method was specifically designed for uniform-sized bursts and required predetermined burst sizes.
Unlike the method proposed in \cite{Shen2019}, we introduce a new hybrid burst-sparse prior that characterizes a more realistic sparse structure without making assumptions about the configuration of the angular bursts.

\begin{figure}
\centering
\includegraphics[width=0.489\textwidth,height=0.152\textwidth]{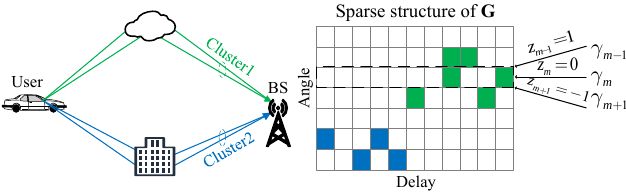}
\caption{Illustration of an MIMO-OTFS system and the corresponding burst and i.i.d. sparse structures of $\G$ within the angular and delay domains.}
\end{figure}

The conventional i.i.d. Gaussian-Gamma prior \cite{Liu2020} is:
\begin{align}\label{i.i.d. prior}
p(\G|\bm\Xi)=\prod_{m=1}^{M_\theta}\prod_{n=1}^{N_{\tau}} \mathcal{CN}(g_{m,n}|0,\xi_{m,n}^{-1}),
\end{align}
where $\mathcal{CN}(\cdot)$ stands for the complex Gaussian distribution, $[\bm\Xi]_{m,n}\triangleq\xi_{m,n}$ is the precision of $g_{m,n}$, directly indicating the support of $\G$, and $[\cdot]_{m,n}$ denotes the $(m,n)$-th element of a matrix.
However, such a simple prior  lacks the capacity to capture both burst and i.i.d. sparse structures.
To exploit the i.i.d. sparsity among the delay domain, we first design
\begin{align}\label{Gaussian-Gamma}
p(\G|\bm\gamma,\bm\rho)=\prod_{n=1}^{N_{\tau}}  \underbrace{\mathcal{CN}\Big(\g_{n}|\mathbf{0},(\rho_n\cdot\mathrm{diag}\{\bm\gamma\})^{-1}\Big)}_{\triangleq p(\g_n|\bm\gamma, \rho_n)},
\end{align}
with $p(\bm\gamma)=\prod_{m=1}^{M_{\theta}}\Gamma(\gamma_m|c,d)$ and $p(\bm\rho)=\prod_{n=1}^{N_{\tau}}\Gamma(\rho_n|c,d)$, where $\bm\rho=[\rho_1,\rho_2,\cdots,\rho_{N{\tau}}]^{\T}$, $\bm\gamma=[\gamma_1,\gamma_2,\cdots,\gamma_{M_{\theta}}]^{\T}$, $\g_n=[g_{1,n},g_{2,n},\cdots,g_{M_\theta,n}]^{\T}$, $\Gamma(\cdot)$ denotes the Gamma distribution and $c,d\rightarrow 0$ for a broad hyperprior \cite{Cao2021}.
It is worth noting that the prior in (\ref{Gaussian-Gamma}) does not depict the actual distribution of the wireless channel; rather, it serves as a sparse-enforcing prior. Here, $\gamma_m$ controls the sparsity of the $m$-th row of $\G$ and the additional variable $\rho_n$ allows the $n$-th element of the
$m$-th row to be zero despite others having significant values.
Then, we introduce the assignment vector $\z=[z_1,z_2,\cdots,z_{M_\theta}]^{\T}$ with $z_m\in\{-1,0,1\}$ to enforce the burst-sparsity of $\g_n$ and handle outliers simultaneously:
\begin{align}
&p(\g_n|\bm\gamma,\rho_n,\z)\notag\\
=&\prod_{m=1}^{M_{\theta}}\prod_{u=-1}^1
\{\mathcal{CN}(g_{m,n}|0,\gamma_{m+u}^{-1}\rho_n^{-1})\}^{\delta(z_{m}-u)},
\end{align}
with $p(\z)=\prod_{m=1}^{M_{\theta}}\big(\frac{1}{3}\big)^{\delta(z_m+1)}\big(\frac{1}{3}\big)^{\delta(z_m)}\big(\frac{1}{3}\big)^{\delta(z_m-1)}$. In the presence of $\z$, the precision $\gamma_m$ is able to  jointly control the sparsity of the $m$-th row and its neighbours. For example, if the $(m-1)$-th (or $(m+1)$-th) row exhibits significant values,  there is a higher likelihood that the $m$-th row will also display a significant value by leveraging  $\gamma_{m-1}$ (or $\gamma_{m+1}$) while setting $z_m=-1$ (or $z_m=1$). If the $m$-th row is an outlier, we can take $z_m=0$ to utilize  its own $\gamma_m$  to account for outlier.

Under the white complex Gaussian noise assumption,
$p(\Y|\G,\alpha;\bm\beta,\bm\kappa)$ is a complex Gaussian distribution with mean $\A(\bm\beta)\G\S^\T(\bm\kappa)$ and elemental precision $\alpha\triangleq \sigma^{-2}$,
and $\alpha$ is also modeled as a Gamma prior $p(\alpha)=\Gamma(\alpha|c,d)$.
Combining all priors,  the joint probability density function (PDF) becomes:
\begin{align}\label{pdf}
p(&\Y,\G,\bm\gamma,\bm\rho,\z,\alpha;\bm\beta,\bm\kappa)\notag\\
&=p(\Y|\G,\alpha;\bm\beta,\bm\kappa)p(\G|\bm\gamma,\bm\rho,\z)p(\z)p(\bm\gamma)p(\bm\rho)p(\alpha).
\end{align}

\subsection{Independent VBI Factorization-Based Inference}
As discussed in Section~II, the vectorization transformation (\ref{SM-model}) is  computationally inefficient.
Hence, we handle the model (\ref{MM-model})  directly to reduce computational complexity by utilizing the innovative independent VBI factorization \cite{Cao2021}:
\begin{align}\label{VBI}
q(\bm{\Theta})=\left\{\prod_{n=1}^{N_{\tau}}q(\g_n)\right\}q(\bm{\gamma})q(\bm{\rho})q(\alpha)q(\z),
\end{align}
where $\bm{\Theta}\triangleq \{\g_1,\g_2,\cdots,\g_{N_{\tau}},\bm{\gamma},\bm{\rho},\alpha,\z\}$.

Under the VBI framework, the optimal chosen $q^\star(\bm\Theta)$ should minimize the Kullback-Leibler (KL) divergence \cite{Tzikas2008}:
\begin{align}\label{KL}
q^\star(\bm\Theta)=\arg\min_{q(\bm\Theta)}\int q(\bm\Theta) \ln \frac{q(\bm\Theta)}{p(\bm\Theta|\Y)} \prod_s d \Theta_s,
\end{align}
where $\Theta_s$ is the $s$-th element of $\bm{\Theta}$.
The optimal factorization $q^\star(\Theta_s)$ should satisfy \cite{Cao2021}:
\begin{align}\label{updating-rule}
\ln \,q^\star(\Theta_s)\propto\left\langle \,\ln \,p(\Y,\bm{\Theta})\,\right\rangle_{\prod_{s^{\prime}\neq s}q^\star(\Theta_{s^{\prime}})},~\forall s,
\end{align}
where $\langle\cdot\rangle_{q(\cdot)}$ denotes the expectation w.r.t $q(\cdot)$.
Following the standard Bayesian inference \cite{Tzikas2008}, we can find a stationary solution of (\ref{updating-rule}) by resorting to an iterative update algorithm:
\begin{align}
&q^{\new}(\g_n)=\mathcal{CN}(\g_n\,|\,\bm{\mu}_n,\bm{\Sigma}_n),~\forall n,\label{update1}\\
&q^{\new}(\bm\gamma)=\prod_{m=1}^{M_\theta}\Gamma(\gamma_m|c_{\gamma_m}, d_{\gamma_m}),\label{update2}\\
&q^{\new}(\bm\rho)=\prod_{n=1}^{N_\tau}\Gamma(\rho_n|c_{\rho_n}, d_{\rho_n}),\label{update3}\\
&q^{\new}(\alpha)=\Gamma(\alpha|c_{\alpha}, d_{\alpha}),\label{update4}\\
&q^{\new}(\z)=\prod_{m=1}^{M_\theta}\prod_{u\in \{-1,0,1\}}\hat z_{m,u}\delta( z_m-u),\label{update5}
\end{align}
where
%\begin{align}
%~~~~~~~~~~~~\bm{\mu}_n&=\hat{\alpha}\bm{\Sigma}_n\A^{\mathrm{H}}(\bm\beta)\Y_{-n}\s_n^{*}(\kappa_n),\label{calculate parameter1}\\
%\bm{\Sigma}_n&=\big(\hat{\alpha}\epsilon_n\A^{\mathrm{H}}(\bm\beta)\A(\bm\beta)+\hat{\rho}_n\bm\Upsilon\big)^{-1},\label{calculate parameter2}\\
%\hat{\gamma}_m&=c_{\gamma_m}/d_{\gamma_m},\label{calculate parameter3}\\
%\hat{\rho}_n&=c_{\rho_n}/d_{\rho_n},\label{calculate parameter4}\\
%\hat{\alpha}&=c_{\alpha}/d_{\alpha},\label{calculate parameter5}\\
%\hat z_{m,u}&= \exp( \phi_{m,u})/\sum\nolimits_{u\in\{-1,0,1\}}\exp(\phi_{m,u}),\label{calculate parameter6}
%\end{align}
$\bm{\mu}_n=\hat{\alpha}\bm{\Sigma}_n\A^{\mathrm{H}}(\bm\beta)\Y_{-n}\s_n^{*}(\kappa_n)$,
$\bm{\Sigma}_n=\big(\hat{\alpha}\epsilon_n\A^{\mathrm{H}}(\bm\beta)\A(\bm\beta)+\hat{\rho}_n\bm\Upsilon\big)^{-1}$,
$\hat{\gamma}_m=c_{\gamma_m}/d_{\gamma_m}$,
$\hat{\rho}_n=c_{\rho_n}/d_{\rho_n}$,
$\hat{\alpha}=c_{\alpha}/d_{\alpha}$,
$\hat z_{m,u}= \exp( \phi_{m,u})/\sum\nolimits_{u\in\{-1,0,1\}}\exp(\phi_{m,u})$,
with
$c_{\gamma_m}=c+N_{\tau}\sum_{u}\hat z_{m-u,u}$,
$d_{\gamma_m}=d+\sum_{n}\hat\rho_n\sum_{u}\hat z_{m-u,u}\varpi_{m-u,n}$,
$c_{\rho_n}=c+\sum_{m}\sum_{u}\hat z_{m,u}$,
$d_{\rho_n}=d+\sum_{m}\sum_{u}\hat z_{m,u}\gamma_{m+u}\varpi_{m,n}$,
$c_\alpha=c+N_{BS}L$,
$d_\alpha= d+\big\|\Y-\A(\bm\beta)\U\S^{\T}(\bm\kappa)\big\|_{\mathrm{F}}^2+\sum_{n} \epsilon_n\tr\big(\A^{\mathrm{H}}(\bm\beta)\A(\bm\beta)\bm{\Sigma}_n\big)$,
$\phi_{m,u}=N_\tau \widehat{\ln\gamma}_{m+u}-\hat\gamma_{m+u}\sum_{n}\hat \rho_n \varpi_{m,n}$,
$\widehat{\ln\gamma}_m=\langle\,\ln\gamma_m\,\rangle_{q^{\new}(\gamma_m)}=\Psi(c_{\gamma_m})-\ln(d_{\gamma_m})$,
$\Y_{-n}\triangleq\Y-\sum_{n^{\prime}\neq n}  \A (\bm\beta)  \bm\mu_{n^{\prime}}   \s_{n^{\prime}}^{\T}(\kappa_{n^{\prime}})$,
$\bm\Upsilon\!\triangleq\!\sum_{u}\!\diag\{\hat z_{1,u},\hat z_{2,u},\!\cdots\!,\hat z_{M_\theta,u}\} \bm\Lambda_u$,
$\bm\Lambda_{-1}\!\triangleq\!\diag  \{ \!\hat \gamma_{M_{\theta}} , \hat \gamma_1,\hat \gamma_2,\cdots,\hat \gamma_{M_{\theta}-1}\}$,
$\bm\Lambda_0\triangleq\diag\{\hat \gamma_1,\hat \gamma_2,\cdots,\hat \gamma_{M_{\theta}}\}$,
$\bm\Lambda_1\triangleq\diag\{\hat \gamma_2,\hat \gamma_3,\cdots,\hat \gamma_{M_{\theta}},\hat \gamma_1\}$,
$\U\triangleq[\bm\mu_1,\bm\mu_2,\cdots,\bm\mu_{N_\tau}]$,
$\varpi_{m,n}\triangleq \big|[\U]_{m,n}\big|^2+[\bm\Sigma_n]_{m,m}$,
$\epsilon_n\triangleq \s^{\mathrm{H}}_n(\kappa_n)\s_n(\kappa_n)$,
$(\cdot)^*$ being conjugate operator, and $\Psi(\cdot)$ being digamma function.

\subsection{Angle-Doppler Refinement}
This subsection focuses on the refinement of off-grid gaps $\bm\beta$ and Doppler-shifts $\bm\kappa$ via the EM algorithm \cite{Yang2013}.
The estimated $\bm{\beta}$ and $\bm\kappa$ should maximize the target distribution (\ref{pdf}), i.e.,
\begin{align}
\!\!\!(\bm{\beta}^{\new},\bm\kappa^{\new})\!=\!\arg\max_{\bm{\beta},\bm\kappa}
\big\langle p(\Y|\G,\alpha;\bm\beta,\bm\kappa)\big\rangle_{\prod_n q(\g_n)q(\alpha)}.
\end{align}
Since the objective function is non-convex w.r.t. $\bm\beta$, we apply the first-order Taylor expansion \cite{Yang2013} as:
$\A(\bm\beta)\approx\A+\B\diag(\bm\beta)$,
where $\B=[\b(\theta_1),\b(\theta_2),\cdots,\b(\theta_{M_\theta})]$ with $\b(\theta_m)$ standing for the derivative of $\a(\theta_m)$.
Setting the derivative of the objective function w.r.t. $\bm\beta$ to zero yields \cite{Yang2013}:
\begin{align}\label{angle-refine}
\bm{\beta}^{\new}=\mathbf{P}^{-1}\v,
\end{align}
where
$\mathbf{P}\triangleq \Re\big\{(\B^{\mathrm{H}} \B)^* \odot \big(\U \mathbf{S}^{\T} (\bm\kappa)\mathbf{S}^{*} (\bm\kappa)\U^{\mathrm{H}}+ \sum_n \epsilon_n \bm{\Sigma}_n\big)\big\}$ and
$\v \triangleq \Re\big\{  \sum_{t=1}^{L} \diag^* (\U\tilde\s_t(\bm\kappa)) \B^{\mathrm{H}} (\tilde\y_t - \A \U \tilde\s_t(\bm\kappa))  \big\}- \Re\big\{\\ \diag \big(\B^{\mathrm{H}} \A \sum_n \epsilon_n\bm{\Sigma}_n\big)     \big  \}$
with
$\Re(\cdot)$ denoting the real part, $\tilde\s_t(\bm\kappa)\triangleq [  [\S(\bm\kappa)]_{t,1},[\S(\bm\kappa)]_{t,2},\cdots,[\S(\bm\kappa)]_{t,N_\tau}]^\T$ and
$\tilde\y_t\triangleq [[\Y]_{1,t},[\Y]_{2,t},\cdots,[\Y]_{N_{BS},t}]^\T$.
Then, we refine $\bm{\theta}$ as \cite{Dai2018}:
\begin{align}\label{grid-refine}
\bm\theta^{\new}=\bm{\theta}+\bm{\beta}^{\new}.
\end{align}
The validation of (\ref{grid-refine}) comes from the fact that $\bm\theta^{\mathrm{new}}$ is much closer to the true AoAs than $\bm{\theta}$. Consequently, the error incurred by the first-order Taylor expansion at $\bm\theta^{\mathrm{new}}$ is much smaller than the error incurred at $\bm{\theta}$. After several iterations, the updated grid $\bm\theta^{\mathrm{new}}$ will tend to approach the true AoAs.
A similar way to update the angular grid $\bm\theta^{\mathrm{new}}$ is through gradient descent resorting to Newton's method \cite{Wan2022,Zhang2021Gridless}, which is able to converge to a fixed solution akin to the proposed method.

Since refining $\kappa_n$ is equivalent to refining $\omega_n\triangleq e^{j2\pi\frac{\kappa_n}{L}}$,
we take the derivative of the objective function w.r.t. $\omega_n$ as:
\begin{align}
\frac{\partial\ln\langle p(\Y|\G,\alpha;\bm\beta,\bm\kappa)\rangle_{\prod_n q(\g_n)q(\alpha)}}{\partial \omega_n}=\sum\nolimits_{t=1}^{L}\varepsilon_{t,n}\omega_n^{t},
\end{align}
where
$\varepsilon_{1,n}\triangleq\sum_{t=1}^{L}(t-1)|[\X]_{t,n}|^2 \varrho_n$,
$\varepsilon_{t,n}\triangleq (t-1)[\X]_{t,n}(-\tilde\y_t^{\mathrm{H}}\A(\bm\beta)\bm\mu_t+\sum_{n^{\prime}\neq n }[\S]_{t,n^{\prime}}^*\varrho_n), \forall t>1$,
$\varrho_n\triangleq \bm\mu_n^{\mathrm{H}}\A^{\mathrm{H}}(\bm\beta)\A(\bm\beta)\bm\mu_n$,
and
$\X\triangleq [\bm\Pi \x,\bm\Pi^2 \x,\cdots,\bm\Pi^{N_{\tau}} \x]$.
Selecting the closest root to the unit circle as $\omega_n^{\new}$, $\kappa_n^{\new}$ becomes:
\begin{align}\label{Doppler-refine}
\kappa_n^{\new}=\frac{L}{2\pi}\angle\,\omega_n^{\new},
\end{align}
where $\angle\,\omega_n^{\new}$ denotes the angle of $\omega_n^{\new}$.

Generally, the proposed method iteratively updates $q(\g_n)$s, $q(\bm\gamma)$, $q(\bm\rho)$, $q(\alpha)$, $q(\z)$, $\bm\theta^{\new}$, and $\kappa_n^{\new}$ using (\ref{update1})-(\ref{update5}), (\ref{grid-refine}) and (\ref{Doppler-refine}), respectively, until convergence. In the following, we provide a brief outline of the convergence proof for the proposed method.
Since the updating rules (\ref{update1})-(\ref{update5}) yield uniquely determined distributions, the functional optimization problem (\ref{KL}) can be transformed into a parameterized optimization problem. It can then be shown that the sequence generated by this parameterized optimization problem converges to a limit. Finally, we establish that this limit is a stationary point. The detailed proof can be found in Appendix~B.
The computational complexity of the proposed method primarily stems from updating $q(\g_n)$s via (\ref{update1}), which is $\mathcal{O}(N_\tau N_{BS}^2 M_\theta)$ per iteration,
 while that of the standard VBI \cite{Tzikas2008} is $\mathcal{O}(L^2 N_{BS}^2 N_{\tau} M_{\theta})$ per iteration.
Thus, the proposed method significantly reduces complexity compared to the standard VBI.
A detailed computational complexity analysis can be found in Appendix~C.

\section{Simulation Validation}

This section conducts several simulations to evaluate the performance of the proposed method using Monte Carlo trials.
The considered OTFS system is equipped with a uniform linear array whose distance between adjacent antennas is $\lambda/2$.
The system employs $M=256$ subcarriers and $N=128$ frames, with a carrier frequency of $f_0=6\,\mathrm{GHz}$ and a subcarrier spacing of $\Delta\!f=15\,\mathrm{kHz}$. We consider a user's speed ranging from $0$ to $\textsl{v}_{\mathrm{u}}=100\,\mathrm{m/s}$ and a carrier frequency of $f_0=6\,\mathrm{GHz}$. This corresponds to a Doppler frequency range of $\pm f_0\textsl{v}_{\mathrm{u}}/\textsl{v}_{\mathrm{c}}=\pm 2000\,\mathrm{Hz}$, where $\textsl{v}_{\mathrm{c}}$ is the speed of light.
The channel is assumed to contain $2$ scattering clusters ranging from $[-60^ {\circ}, 60^ {\circ}]$.
Each cluster has $10$ sub-paths, concentrated within a $\pm 3^ {\circ}$ angle spread.
The delays and Doppler-shifts are uniformly chosen from $[\frac{1}{M\Delta\!f},\frac{2}{M\Delta\!f}, \cdots, \frac{20}{M\Delta\!f}]$ and $[-2000\, \mathrm{Hz}, 2000\, \mathrm{Hz}]$, respectively.
The angle grid uniformly covers $[-90^ {\circ}, 90^ {\circ}]$ with $M_\theta = 90$, and the delay grid lies on $[\frac{1}{M\Delta\!f},\frac{2}{M\Delta\!f}, \cdots, \frac{N_\tau}{M\Delta\!f}]$, where $N_\tau=20$.
The channel model $\H$ is derived from (\ref{input-output}) as $\y\triangleq\mathrm{vec}(\Y)=\H\x+\w$,
where $\H=\sum_{p}\xi_p(\bm\Delta^{k_p}\bm\Pi^{l_p})\otimes\a(\vartheta_p)$. We evaluate the performance of the proposed method using the normalized mean square error (NMSE) of $\H$, defined as:
$\frac{1}{I_t} \sum_{i_t=1}^{I_t} \| \H_{i_t} - \hat{\H}_{i_t} \|_{\mathrm{F}}^2/\| \H_{i_t}  \|_{\mathrm{F}}^2$,
where $\hat{\H}_{i_t}$ is the estimation of $\H_{i_t}$ at the $i_t$-th Monte Carlo trial, and the maximum number of trials is $I_t=200$.
\begin{figure}
\center
\begin{tikzpicture}
\begin{semilogyaxis}[xlabel={  SNR [dB]},
ylabel={ {NMSE}},
title={\large (a)},
grid=major,
legend style={at={(0.695,1.000)},
anchor=north,legend columns=2},xmin=-10,xmax=20,ymin=0.001,ymax=10]
\addplot[mark=square,black]  coordinates{(-10, 9.32405) (-5, 3.03627) (0, 1.018719) (5, 0.32718) (10, 0.1027799) (15, 0.0316013) (20, 0.01082)};
\addlegendentry{\scriptsize LS}
\addplot[mark=x,black]  coordinates{(-10, 8.32405) (-5, 2.63627) (0, 0.818719) (5, 0.26718) (10, 0.0827799) (15, 0.0275911) (20, 0.00924504)};
\addlegendentry{\scriptsize $\ell_1$-norm}
\addplot[mark=+,black]  coordinates { ( -10, 4.59358 ) ( -5, 1.52309 ) ( 0, 0.478714 ) ( 5, 0.158888 ) ( 10, 0.0507708 ) ( 15, 0.0179936 ) ( 20, 0.00634056 ) };
\addlegendentry{\scriptsize OGVBI}
\addplot[mark=o,black]  coordinates{(-10, 2.07833) (-5, 0.674053) (0, 0.218903) (5, 0.0720025) (10, 0.026856) (15, 0.00922164) (20, 0.0036291) };
\addlegendentry{\scriptsize Vector-OGVBI}
\addplot[mark=triangle,black]  coordinates{(-10, 1.93772) (-5, 0.625126) (0, 0.199584) (5, 0.0640737) (10, 0.0228457) (15, 0.00782822) (20, 0.00304913) };
\addlegendentry{\scriptsize Fast-VBI}
\addplot[mark=diamond,red]  coordinates{(-10, 0.41022) (-5, 0.170) (0, 0.063) (5, 0.0251283) (10, 0.0101) (15, 0.00403233) (20, 0.00165)};
\addlegendentry{\scriptsize Proposed}
\end{semilogyaxis}
\end{tikzpicture}
\begin{tikzpicture}
\begin{semilogyaxis}[xlabel={ $L$  },
ylabel={  {NMSE}},
title={\large (b)},
grid=major,
xmin=20,xmax=100,ymin=0.002,ymax=1]
\addplot[mark=square,black]  coordinates{(20, 0.823087125208597) (30, 0.131772229496632) (40, 0.0822852370379394) (50, 0.0592429603283734) (60, 0.0481869978760306) (70, 0.0377756387107733) (80, 0.0344436751875461) (90, 0.0310447305045702) (100, 0.0246657677349904)};								
\addplot[mark=x,black]  coordinates{(20, 0.608375724233054) (30, 0.106641047268278) (40, 0.0750454266682596) (50, 0.0537303772526387) (60, 0.0430569214665662) (70, 0.0350833323045012) (80, 0.0319393646887337) (90, 0.0292348799005075) (100, 0.0231672493031695)};																
\addplot[mark=+,black]  coordinates{(20, 0.419254589650043) (30, 0.0770566105690097) (40, 0.0470466804203386) (50, 0.0342078974804508) (60, 0.0273681113676352) (70, 0.0227541173004974) (80, 0.0208983702295043) (90, 0.0195024516782652) (100, 0.0184649060429205)};
\addplot[mark=o,black]  coordinates{(20, 0.0718710132995838) (30, 0.0334402012704830) (40, 0.0220725586016768) (50, 0.0171656235999946) (60, 0.0142774803830480) (70, 0.0122114378514255) (80, 0.0109135933017961) (90, 0.00999021179870043) (100, 0.00933577678182254)};																
\addplot[mark=triangle,black]  coordinates{(20, 0.0370814731267500) (30, 0.0250826088023420) (40, 0.0182131686494545) (50, 0.0146662367641058) (60, 0.0124748627312033) (70, 0.0109762389514563) (80, 0.00976194918445251) (90, 0.00899257373566748) (100, 0.00846236203513440)};								
\addplot[mark=diamond,red]  coordinates{(20, 0.0252343967368708) (30, 0.0139533713701027) (40, 0.00971772666727662) (50, 0.00797288232772952) (60, 0.00678877434472127) (70, 0.00597098098365511) (80, 0.00545493079079692) (90, 0.00531831339130338) (100, 0.00514906635159068)};								
\end{semilogyaxis}
\end{tikzpicture}
\begin{tikzpicture}
\begin{semilogyaxis}[xlabel={ $N_{BS}$  },
ylabel={  {NMSE}},
title={\large (c)},
grid=major,
xmin=10,xmax=80,ymin=0.009,ymax=0.16]
\addplot[mark=square,black]  coordinates{(10, 0.14241)	(20, 0.1420)	(30, 0.1456)	(40, 0.1426)	(50, 0.1445)	(60, 0.1509)	
                                    (70, 0.15098)	(80, 0.1453)};
\addplot[mark=x,black]  coordinates{(10, 0.131713729006037)	(20, 0.132380930651561)	(30, 0.134878105238861)	(40, 0.132640293123376)	(50, 0.134001916860195)	(60, 0.141604254526674)	
                                    (70, 0.134999163308940)	(80, 0.136869139372684)};
\addplot[mark=+,black]  coordinates{(10, 0.0935810054827654) (20, 0.0806263650485953) (30, 0.0720383508000541) (40, 0.0641084505000937)	(50, 0.0603821954504237) (60, 0.0531554821876599)
                                    (70, 0.0486292601864668) (80, 0.0409038970213248)};
\addplot[mark=o,black]  coordinates{(10, 0.0797527140099268)	(20, 0.0517027742219184) (30, 0.0440509905482408) (40, 0.0390777064270125) (50, 0.0366892931765447)	(60, 0.0348821838640596)
                                    (70, 0.0333537834911563)	(80, 0.0314985812148376)};
\addplot[mark=triangle,black]  coordinates{(10, 0.0590189529639814)	(20, 0.0416773442908458)	(30, 0.0363501398706936)	(40, 0.0323911941546709)	(50, 0.0304229918158696)	(60, 0.0285851472676971)	(70, 0.0273396344402125)	(80, 0.0254979307632755)};
\addplot[mark=diamond,red]  coordinates{(10, 0.0369650596189836)	(20, 0.0230714071683783)	(30, 0.0170785369558124)	(40, 0.0144139685488258)	(50, 0.0125763655412150)	(60, 0.0112544249005498) (70, 0.0102494127067176)	(80, 0.00977106431780082)};							
\end{semilogyaxis}
\end{tikzpicture}
\caption{NMSEs of uplink OTFS channel estimation under different simulation scenarios. (a) NMSE versus SNR with $L=40$ and $N_{BS}=40$; (b) NMSE versus pilot number $L$ with $\mathrm{SNR}=10\,\dB$ and $N_{BS}=40$; (c) NMSE versus antenna number $N_{BS}$ with $L=30$ and $\mathrm{SNR}=10\,\dB$.}
\label{Simulation}
\end{figure}
\begin{figure}
\center
\begin{tikzpicture}
\begin{semilogyaxis}[xlabel={ Number of delay grid point $N_\tau$ },
ylabel={ Runtime [second]},
title={},
grid=major,
legend style={at={(0.70,1)},
anchor=north,legend columns=2},xmin=20,xmax=40,ymin=0.01,ymax=250]
\addplot[mark=square,black]  coordinates{(20, 3.91352) (25, 7.89453) (30, 13.221) (35, 20.9301) (40, 31.961)};
\addlegendentry{\scriptsize LS}
\addplot[mark=x,black]  coordinates{(20, 3.20092) (25, 3.21764) (30, 3.1834) (35, 3.37656) (40, 3.46658)};
\addlegendentry{\scriptsize $\ell_1$-norm}
\addplot[mark=+,black]  coordinates { (20, 0.0619657) (25, 0.0582868) (30, 0.0581051) (35, 0.057268 ) (40, 0.0554986)};
\addlegendentry{\scriptsize OGVBI}
\addplot[mark=o,black]  coordinates{(20, 27.2028) (25, 60.4179) (30, 106.214) (35, 143.676) (40, 211.4)};
\addlegendentry{\scriptsize  Vector-OGVBI}
\addplot[mark=triangle,black]  coordinates{(20, 1.32249) (25, 1.40401) (30, 1.45097) (35, 1.59853) (40, 1.68322)};
\addlegendentry{\scriptsize Fast-VBI}
\addplot[mark=diamond,red]  coordinates{(20, 1.6818) (25, 1.66382) (30, 1.86208) (35, 1.88396) (40, 1.94523)};
\addlegendentry{\scriptsize Proposed}

\end{semilogyaxis}
\end{tikzpicture}
\caption{Computational time versus the number of delay grid points $N_\tau$ with $L=40$, $N_{BS}=40$ and $M_\theta=90$.}
\label{time}
\end{figure}

%Simulation~1 aims to systematically investigate the impact of independent VBI factorization and angle-Doppler refinement on NMSE performance. We conduct a comparative analysis of the proposed method against three ablation variants:
%\begin{itemize}
%  \item\emph{Variant\,1:} Replace the independent VBI factorization (\ref{VBI}) with the standard VBI factorization in the proposed method.
%  \item\emph{Variant\,2:} Omit angle-Doppler refinement (\ref{angle-refine}) and (\ref{Doppler-refine}) in the proposed method.
%  \item\emph{Variant\,3:} Replace  the independent VBI factorization (\ref{VBI}) with the standard VBI factorization in the proposed method, and omit angle-Doppler refinement (\ref{angle-refine}) and (\ref{Doppler-refine}).
%\end{itemize}
%Fig.~2 shows the NMSEs (versus SNR) of different ablation variants, with $L=40$ and $N_{BS}=40$.
%It reveals that:
%(i) Variant~1 achieves the best NMSE performance by employing the standard VBI factorization, but it leads to a significant increase in computational complexity;
%(ii) Variant~3 exhibits a notable NMSE performance degradation compared to Variant~1, due to the omission of angle-Doppler refinement;
%(iii) Variant~3 and Variant~2 show comparable NMSE performance, highlighting the effectiveness of the independent VBI factorization in OTFS channel estimation;
%and (iv) the proposed method strikes a balance between NMSE performance and simplicity, as its NMSE is similar to that of Variant~1, while significantly reducing the computational burden.

Simulation~1 investigates the influence of SNR on the NMSE of different baselines:
(i) LS: recover $\g$ from (\ref{SM-model}) using least squares (LS), where $\bm{\beta}=\bm{0}$ and $\bm{\kappa}$ is given as the true value;
(ii) $\ell_1$-norm: recover $\g$ from (\ref{SM-model}) using $\ell_1$-norm, where $\bm{\beta}=\bm{0}$ and $\bm{\kappa}$ is given as the true value;
(iii) OGVBI: recover $\G$ from (\ref{MM-model}) via off-grid VBI \cite{Yang2013}, where the term $\G\S^{\T}(\bm\kappa)$ in (\ref{MM-model}) is treated as an unknown row-sparse matrix and $\A(\bm{\beta})$ is approximated using the first-order Taylor expansion;
(iv) Vector-OGVBI: recover $\g$ from (\ref{SM-model}) using OGVBI \cite{Liu2020}, while approximating $\A(\bm{\beta})$ and $\S(\bm{\kappa})$ using the first-order Taylor expansion.
and (v) Fast-VBI: recover $\G$ from (\ref{MM-model}) using Fast-VBI \cite{Cao2021}, which is a degenerate version of the proposed method using the i.i.d. prior (\ref{i.i.d. prior}).
With $L=40$ and $N_{BS}=40$, the NMSEs versus SNR shown in Fig.~2(a) reveal that:
(i) the NMSEs of all  the methods decrease as SNR increases;
(ii) compared to LS, $\ell_1$-norm shows improved performance by leveraging angle-delay domain sparsity;
(iii) compared to LS and $\ell_1$-norm, Vector-OGVBI and Fast-VBI show performance improvements due to angle refinement; and
(iv) the proposed method exhibits substantially improved performance due to its superior capability in capturing burst sparsity.

Simulation~2 studies how the pilot length $L$ affects NMSE performance, with $\mathrm{SNR}=10\,\mathrm{dB}$ and $N_{BS}=40$. The NMSE results shown in Fig.~2(b) indicate that:
(i) the NMSE performance of all the methods improves as $L$ increases; and
(ii) Fast-VBI exhibits better performance compared to Vector-OGVBI by eliminating the first-order Taylor approximation.

Simulation~3 examines the impact of $N_{BS}$ on NMSEs, with $\mathrm{SNR}=10\,\mathrm{dB}$ and $L=30$. The NMSE results shown in Fig.~2(c) reveal that:
(i) as $N_{BS}$ increases, the NMSEs of almost all methods decrease, except for LS and $\ell_1$-norm, whose channel estimation accuracy for angular burst-sparse channels is primarily affected by off-grid errors, and therefore increasing $N_{BS}$ has a limited effect on their NMSE performance; and
(ii) the performance gaps between the proposed method and others progressively widen as the antenna number $N_{BS}$ increases, owing to the enhanced angle resolution.

Simulation~4 examines the computational time of different strategies versus the number of delay grid points $N_\tau$.
Fig.~3 displays the CPU runtime of different strategies with $L=40$, $N_{BS}=40$ and $M_\theta=90$.
The maximum iterative number for the
proposed method is selected as 80, and that for other methods are selected as 60, based on the
convergence iterative numbers illustrated in Appendix~D.
It is evident that:
(i) unlike Vector-OGVBI, which employs the conventional VBI factorization, both the proposed method and Fast-VBI achieve a notable decrease in computational time by adopting the independent VBI factorization;
(ii) the proposed method boasts lower computational time compared to $\ell_1$-norm and LS-based methods, and these latter two methods are unable to capitalize on channel sparsity to enhance their performance;
and
(iii) although OGVBI exhibits the lowest computational complexity, it fails to harness the sparsity present in the delay domain, leading to a significant decline in NMSE performance.

Simulation~5 provides a comparison of the sparse enforcing of the proposed hybrid burst sparse prior with i.i.d. sparse prior, where two clusters are considered and each cluster has 5 sub-paths. The true AoAs of clusters are in two burst blocks of $\{-10^ {\circ}, -9^ {\circ}, -8^ {\circ}, -7^ {\circ}, -6^ {\circ}\}$ and $\{10^ {\circ}, 11^ {\circ}, 12^ {\circ}, 13^ {\circ}, 14^ {\circ}\}$, respectively. Setting $\mathrm{SNR}=5\,\mathrm{dB}$, pilot length $L=40$ and antenna number $N_{BS}=40$, Fig.~4 shows the recovered sparse channel in the angular domain using the proposed hybrid burst sparse prior and the conventional i.i.d. prior, respectively. It is observed that:
(i) the conventional i.i.d. prior leads to a leakage of energy over some random positions, which results in a serious performance loss because the random positions are usually far from the true positions;
and (ii) the proposed hybrid burst sparse prior can greatly improve the sparsity and accuracy of the sparse channel recovery, as the energy converges to the true angular bursty blocks.
\begin{figure}
\center
\begin{tikzpicture}[scale=1]
\begin{axis}
[xlabel={ AoA [degree]},
ylabel={Modulus},
legend style={at={(0.28,0.83),
font=\footnotesize},
anchor=north,legend columns=1
}, xmin=-30,ymin=0,xmax=30,ymax=21,title={\large (a)}]
\addplot+[ycomb,color=blue,mark=o] file {Burst_for_our.txt} ;
\addplot[style= dashed] coordinates { (-10,0) (-10,21) };
\addplot[style= dashed] coordinates { (-9,0) (-9,21) };
\addplot[style= dashed] coordinates { (-8,0) (-8,21) };
\addplot[style= dashed] coordinates { (-7,0) (-7,21) };
\addplot[style= dashed] coordinates { (-6,0) (-6,21) };
\addplot[style= dashed] coordinates {  (10,0)  (10,21) };
\addplot[style= dashed] coordinates {  (11,0)  (11,21) };
\addplot[style= dashed] coordinates {  (12,0)  (12,21) };
\addplot[style= dashed] coordinates {  (13,0)  (13,21) };
\addplot[style= dashed] coordinates {  (14,0)  (14,21) };
%\node (a) at (axis cs:36.5,4.7) {\large $\hat L = 200$ };
%\legend{Our method, True DOAs}
\end{axis}
\end{tikzpicture}
\begin{tikzpicture}[scale=1]
\begin{axis}
[xlabel={ AoA [degree]},
ylabel={Modulus},
legend style={at={(0.28,0.83),
font=\footnotesize},
anchor=north,legend columns=1
}, xmin=-30,ymin=0,xmax=30,ymax=21,title={\large (b)}]
\addplot+[ycomb,color=blue,mark=o] file {Burst_for_iid.txt} ;
\addplot[style= dashed] coordinates { (-10,0) (-10,21) };
\addplot[style= dashed] coordinates { (-9,0) (-9,21) };
\addplot[style= dashed] coordinates { (-8,0) (-8,21) };
\addplot[style= dashed] coordinates { (-7,0) (-7,21) };
\addplot[style= dashed] coordinates { (-6,0) (-6,21) };
\addplot[style= dashed] coordinates {  (10,0)  (10,21) };
\addplot[style= dashed] coordinates {  (11,0)  (11,21) };
\addplot[style= dashed] coordinates {  (12,0)  (12,21) };
\addplot[style= dashed] coordinates {  (13,0)  (13,21) };
\addplot[style= dashed] coordinates {  (14,0)  (14,21) };
%\node (a) at (axis cs:36.5,4.7) {\large $\hat L = 200$ };
%\legend{Our method, True DOAs}
\end{axis}
\end{tikzpicture}
\caption{Modulus of the recovered sparse channel in angular domain with $\mathrm{SNR}=5\,\mathrm{dB}$, pilot length $L=40$ and antenna number $N_{BS}=40$. The true AoAs are denoted by dotted lines. (a) Use the proposed hybrid burst sparse prior. (b) Use the conventional i.i.d. prior.}
\end{figure}

%\begin{figure}
%\centering
%\includegraphics[scale=1]{a1}
%
%~~~~~(a)
%
%\includegraphics[scale=1]{a2}
%
%~~~~~(b)
%
%\caption{Modulus of the recovered sparse channel in angular domain with $\mathrm{SNR}=5\,\mathrm{dB}$, pilot length $L=40$ and antenna number $N_{BS}=40$. The true AoAs are denoted by dotted lines. (a) Use the proposed hybrid burst sparse prior. (b) Use the conventional i.i.d. prior.}
%\end{figure}

Simulation~6 demonstrates that the burst sparse prior performs well even when the channel does not exhibit burst sparsity in an experimental trial. Consider an example where the true AoAs of burst-sparse channel are in a burst block of $\{-10^ {\circ}, -9^ {\circ}, -8^ {\circ}, -7^ {\circ}, -6^ {\circ}\}$, and the true AoAs of i.i.d. sparse channel are three individual values of $5^ {\circ}$, $16^ {\circ}$ and $19^ {\circ}$. Fig.~5 shows the recovered sparse channel in the angular domain using the proposed hybrid burst sparse prior, where $\mathrm{SNR}=20\,\mathrm{dB}$, the pilot length is $L=40$ and the antenna number is  $N_{BS}=40$. It is observed that the proposed hybrid burst sparse prior accurately captures all the true AoAs, whether within burst blocks or not, verifying that it performs well for both burst-sparse and classical i.i.d. sparse channels.
%\begin{figure}
%\centering
%\includegraphics[scale=0.75]{b}
%\caption{Modulus of recovered sparse channel in angular domain with $\mathrm{SNR}=20\,\mathrm{dB}$, pilot length $L=40$ and antenna number $N_{BS}=40$. The true AoAs are denoted by dotted lines.}
%\end{figure}

\begin{figure}
\center
\begin{tikzpicture}[scale=1]
\begin{axis}
[xlabel={AoA [degree]},
ylabel={Modulus},
legend style={at={(0.28,0.83),
font=\footnotesize},
anchor=north,legend columns=1
}, xmin=-30,ymin=0,xmax=30,ymax=21,title={}]
\addplot+[ycomb,color=blue,mark=o] file {Burst_iid.txt} ;
\addplot[style= dashed] coordinates { (-10,0) (-10,21) };
\addplot[style= dashed] coordinates { (-9,0) (-9,21) };
\addplot[style= dashed] coordinates { (-8,0) (-8,21) };
\addplot[style= dashed] coordinates { (-7,0) (-7,21) };
\addplot[style= dashed] coordinates { (-6,0) (-6,21) };
\addplot[style= dashed] coordinates {  (5,0)  (5,21) };
\addplot[style= dashed] coordinates {  (16,0)  (16,21) };
\addplot[style= dashed] coordinates {  (19,0)  (19,21) };

%\node (a) at (axis cs:36.5,4.7) {\large $\hat L = 200$ };
%\legend{Our method, True DOAs}
\end{axis}
\end{tikzpicture}

\caption{Modulus of recovered sparse channel in angular domain with $\mathrm{SNR}=20\,\mathrm{dB}$, pilot length $L=40$ and antenna number $N_{BS}=40$. The true AoAs are denoted by dotted lines.}
\end{figure}
\section{Conclusion}
This letter addresses the challenge of MIMO-OTFS channel estimation by introducing a novel burst-sparse prior designed to capture both burst and i.i.d. sparsity in the angular and delay domains, respectively. Moreover, we utilize an independent VBI factorization integrated with angle-Doppler refinement, significantly reducing computational complexity and enhancing estimation performance.

\appendices
\section{}
This appendix compares the maximum phase accumulations caused by Doppler-shifts, angles, and delays, respectively, demonstrating that treating Doppler-shifts as parameters results in a slightly smaller performance loss for OTFS channel estimation.
Consider an OTFS system equipped with a uniform linear array consisting of $N_{BS}=40$ antennas, operating at a carrier frequency $f_0=6 \,\mathrm{GHz}$ with a subcarrier spacing $\Delta\!f=15\,\mathrm{kHz}$. The system has $M=256$ subcarriers and employs a pilot of $L=40$ samples.
If a user has a speed of $\textsl{v}_{\mathrm{u}}=300\,\mathrm{km/h}$,  its maximum Doppler-shift will be:
\begin{align}
\nu_{\max}=\pm f_0\frac{\textsl{v}_{\mathrm{u}}}{\textsl{v}_{\mathrm{c}}}\approx \pm1666\,\mathrm{Hz},
\end{align}
where $\textsl{v}_{\mathrm{c}}$ is the speed of light. Then, the corresponding maximum phase accumulation can be calculated as:
\begin{align}\label{Doppler-phase}
\phi_{\nu_{\max}}=2\pi\nu_{\max}L\frac{1}{M\Delta\!f}\approx\pm 0.086\pi.
\end{align}
On the other hand, the phase accumulation caused by the angles can be calculated as:
\begin{align}
\phi_{\theta_{\max}}=2\pi N_{BS}d\frac{\sin(\theta_{\max})}{\textsl{v}_{\mathrm{c}}/f_0}=\pm 40\pi,
\end{align}
with $\theta_{\max}=\pm \frac{\pi}{2}$ and $d=\frac{\textsl{v}_{\mathrm{c}}}{2f_0}$; while  the phase accumulation caused by the delays can be calculated as:
\begin{align}
\phi_{\tau_{\max}}=-2\pi f_0\tau_{\max}=-2\pi f_0N_\tau\frac{1}{M\Delta\!f}=-62500\pi,
\end{align}
where the maximum delay tap $N_\tau$ is usually set to  $20$ as in \cite{Liu2020,Raviteja2019,Zhao2020}.
Given that the absolute value of $\phi_{\nu_{\max}}$ is much smaller than that of $\phi_{\theta_{\max}}$ and $\phi_{\tau_{\max}}$, it is reasonable to treat the narrow Doppler domain parametrically while representing the broad angle-delay domain in a sparse manner.

\section{}
In this appendix, we prove that the  proposed OTFS channel estimation algorithm with
independent VBI factorization converges to a stationary solution. As shown in \cite{Dai2019},  the optimization problem (\ref{updating-rule}) can be iteratively solved by:
\begin{align}
&\ln q^{(i+1)}(\g_1)\notag\\
&\propto\left\langle \ln p(\Y,\bm{\Theta})\right\rangle_{q^{(i)}_{\g_2}q^{(i)}_{\g_3}\ldots q^{(i)}_{\g_{N_\tau}}q^{(i)}_{\bm\gamma}q^{(i)}_{\bm\rho}q^{(i)}_{\alpha}q^{(i)}_{\z}},\label{updating1}\\
&\ln q^{(i+1)}(\g_2)\notag\\
&\propto\left\langle \ln p(\Y,\bm{\Theta})\right\rangle_{q^{(i+1)}_{\g_1}q^{(i)}_{\g_3}\ldots q^{(i)}_{\g_{N_\tau}} q^{(i)}_{\bm\gamma}q^{(i)}_{\bm\rho}q^{(i)}_{\alpha}q^{(i)}_{\z}},\label{updating2}\\
&~\,\vdots\notag\\
&\ln q^{(i+1)}(\g_{N_\tau})\notag\\
&\propto\left\langle \ln p(\Y,\bm{\Theta})\right\rangle_{q^{(i+1)}_{\g_1}q^{(i+1)}_{\g_2}\ldots q^{(i+1)}_{\g_{N_\tau-1}} q^{(i)}_{\bm\gamma}q^{(i)}_{\bm\rho}q^{(i)}_{\alpha}q^{(i)}_{\z}},\label{updating3}\\
&\ln q^{(i+1)}(\bm\gamma)\notag\\
&\propto\left\langle \ln p(\Y,\bm{\Theta})\right\rangle_{q^{(i+1)}_{\g_1}q^{(i+1)}_{\g_2}\ldots q^{(i+1)}_{\g_{N_\tau}}q^{(i)}_{\bm\rho}q^{(i)}_{\alpha}q^{(i)}_{\z}},\label{updating4}\\
&\ln q^{(i+1)}(\bm\rho)\notag\\
&\propto\left\langle \ln p(\Y,\bm{\Theta})\right\rangle_{q^{(i+1)}_{\g_1}q^{(i+1)}_{\g_2}\ldots q^{(i+1)}_{\g_{N_\tau}}q^{(i+1)}_{\bm\gamma}q^{(i)}_{\alpha}q^{(i)}_{\z}},\label{updating5}\\
&\ln q^{(i+1)}(\alpha)\notag\\
&\propto\left\langle \ln p(\Y,\bm{\Theta})\right\rangle_{q^{(i+1)}_{\g_1}q^{(i+1)}_{\g_2}\ldots q^{(i+1)}_{\g_{N_\tau}}q^{(i+1)}_{\bm\gamma}q^{(i+1)}_{\bm\rho}q^{(i)}_{\z}},\label{updating6}\\
&\ln q^{(i+1)}(\z)\notag\\
&\propto\left\langle \ln p(\Y,\bm{\Theta})\right\rangle_{q^{(i+1)}_{\g_1}q^{(i+1)}_{\g_2}\ldots q^{(i+1)}_{\g_{N_\tau}}q^{(i+1)}_{\bm\gamma}q^{(i+1)}_{\bm\rho}q^{(i+1)}_{\alpha}},\label{updating7}
\end{align}
where $(\cdot)^{(i)}$ stands for the $i$-th iteration and $q_{(\cdot)}$ is short for $q(\cdot)$, e.g., $q_{\g_1}\triangleq q(\g_1)$.
Following the standard VBI \cite{Tzikas2008}, it is clear that updating rules (\ref{update1})-(\ref{update5}) yield uniquely determined distributions:
$q(\g_1)$, $q(\g_2)$,\,\ldots, and $q(\g_{N_\tau})$ are Gaussian distributions parameterized by $\Omega_{\g_1}\triangleq\{\bm\mu_1,\bm\Sigma_1\}$, $\Omega_{\g_2}\triangleq\{\bm\mu_2,\bm\Sigma_2\}$,\,\ldots, and $\Omega_{\g_{N_\tau}}\triangleq\{\bm\mu_{N_\tau},\bm\Sigma_{N_\tau}\}$, respectively;
$q(\bm\gamma)$, $q(\bm\rho)$ and $q(\alpha)$ are gamma distributions parameterized by $\Omega_{\bm\gamma}\triangleq \{c_{\gamma_m},d_{\gamma_m}\}_{m=1}^{M_\theta}$, $\Omega_{\bm\rho}\triangleq \{c_{\rho_n},d_{\rho_n}\}_{n=1}^{N_\tau}$ and $\Omega_{\alpha}\triangleq \{c_{\alpha},d_{\alpha}\}$, respectively;
$q(\z)$ is a categorical distribution parameterized by $\Omega_{\z}\triangleq \{\phi_{m,u}\}_{m=1,u=-1}^{M_\theta,1}$.
Therefore, the functional optimization problem (\ref{KL}) can be converted into a parameterized optimization problem as follows:
\begin{align}
&\big\{\{\Omega_{\g_n}^\star\}_{n=1}^{N_\tau},\Omega_{\bm\gamma}^\star,\Omega_{\bm\rho}^\star,\Omega_{\alpha}^\star,\Omega_{\z}^\star\big\}\notag\\
&=\arg\min_{\{\Omega_{\g_n}\}_{n=1}^{N_\tau},\Omega_{\bm\gamma},\Omega_{\bm\rho},\Omega_{\alpha},\Omega_{\z}}
\mathcal{U}(\{\Omega_{\g_n}\}_{n=1}^{N_\tau},\Omega_{\bm\gamma},\Omega_{\bm\rho},\Omega_{\alpha},\Omega_{\z}),
\end{align}
where the definition of $\mathcal{U}(\{\Omega_{\g_n}\}_{n=1}^{N_\tau},\Omega_{\bm\gamma},\Omega_{\bm\rho},\Omega_{\alpha},\Omega_{\z})$ can be found in (\ref{KL}).
Then, (\ref{update1})-(\ref{update5}) can be rewritten as:
\begin{align}
\Omega_{\g_1}^{(i+1)}&=\arg\min_{\Omega_{\g_1}}
\mathcal{U}(\Omega_{\g_1},\Omega_{\g_2}^{(i)},\Omega_{\g_3}^{(i)}\ldots,\Omega_{\g_{N_\tau}-1}^{(i)},\notag\\
&~~~~~~~~~~~~~~~~\,\Omega_{\g_{N_\tau}}^{(i)},\Omega_{\bm\gamma}^{(i)},\Omega_{\bm\rho}^{(i)},\Omega_{\alpha}^{(i)},\Omega_{\z}^{(i)}),\label{parameter1}\\
\Omega_{\g_2}^{(i+1)}&=\arg\min_{\Omega_{\g_2}}
\mathcal{U}(\Omega_{\g_1}^{(i+1)},\Omega_{\g_2},\Omega_{\g_3}^{(i)},\ldots,\Omega_{\g_{N_\tau}-1}^{(i)},\notag\\
&~~~~~~~~~~~~~~~~\,\Omega_{\g_{N_\tau}}^{(i)},\Omega_{\bm\gamma}^{(i)},\Omega_{\bm\rho}^{(i)},\Omega_{\alpha}^{(i)},\Omega_{\z}^{(i)}),\label{parameter2}\\
&~\,\vdots\notag\\
\Omega_{\g_{N_\tau}}^{(i+1)}&=\arg\min_{\Omega_{\g_{N_\tau}}}
\mathcal{U}(\Omega_{\g_1}^{(i+1)},\Omega_{\g_2}^{(i+1)},\Omega_{\g_3}^{(i+1)},\ldots,\Omega_{\g_{N_\tau}-1}^{(i+1)},\notag\\
&~~~~~~~~~~~~~~~~\,\Omega_{\g_{N_\tau}},\Omega_{\bm\gamma}^{(i)},\Omega_{\bm\rho}^{(i)},\Omega_{\alpha}^{(i)},\Omega_{\z}^{(i)}),\label{parameter3}
\end{align}
\begin{align}
\Omega_{\bm\gamma}^{(i+1)}&=\arg\min_{\bm\gamma}
\mathcal{U}(\Omega_{\g_1}^{(i+1)},\Omega_{\g_2}^{(i+1)},\Omega_{\g_3}^{(i+1)},\ldots,\Omega_{\g_{N_\tau}-1}^{(i+1)},\notag\\
&~~~~~~~~~~~~~~~~\,\Omega_{\g_{N_\tau}}^{(i+1)},\Omega_{\bm\gamma},\Omega_{\bm\rho}^{(i)},\Omega_{\alpha}^{(i)},\Omega_{\z}^{(i)}),\label{parameter4}\\
\Omega_{\bm\rho}^{(i+1)}&=\arg\min_{\Omega_{\bm\rho}}
\mathcal{U}(\Omega_{\g_1}^{(i+1)},\Omega_{\g_2}^{(i+1)},\Omega_{\g_3}^{(i+1)},\ldots,\Omega_{\g_{N_\tau}-1}^{(i+1)},\notag\\
&~~~~~~~~~~~~~~~~\,\Omega_{\g_{N_\tau}}^{(i+1)},\Omega_{\bm\gamma}^{(i+1)},\Omega_{\bm\rho},\Omega_{\alpha}^{(i)},\Omega_{\z}^{(i)}),\label{parameter5}\\
\Omega_{\alpha}^{(i+1)}&=\arg\min_{\Omega_{\alpha}}
\mathcal{U}(\Omega_{\g_1}^{(i+1)},\Omega_{\g_2}^{(i+1)},\Omega_{\g_3}^{(i+1)},\ldots,\Omega_{\g_{N_\tau}-1}^{(i+1)},\notag\\
&~~~~~~~~~~~~~~~~\,\Omega_{\g_{N_\tau}}^{(i+1)},\Omega_{\bm\gamma}^{(i+1)},\Omega_{\bm\rho}^{(i+1)},\Omega_{\alpha},\Omega_{\z}^{(i)}),\label{parameter6}\\
\Omega_{\z}^{(i+1)}&=\arg\min_{\Omega_{\z}}
\mathcal{U}(\Omega_{\g_1}^{(i+1)},\Omega_{\g_2}^{(i+1)},\Omega_{\g_3}^{(i+1)},\ldots,\Omega_{\g_{N_\tau}-1}^{(i+1)},\notag\\
&~~~~~~~~~~~~~~~~\,\Omega_{\g_{N_\tau}}^{(i+1)},\Omega_{\bm\gamma}^{(i+1)},\Omega_{\bm\rho}^{(i+1)},\Omega_{\alpha}^{(i+1)},\Omega_{\z}).\label{parameter7}
\end{align}
(\ref{parameter1})-(\ref{parameter7}) guarantee that the objective function $\mathcal{U}(\{\Omega_{\g_n}\}_{n=1}^{N_\tau},\Omega_{\bm\gamma},\Omega_{\bm\rho},\Omega_{\alpha},\Omega_{\z})$ is iteratively nonincreasing. Since $\mathcal{U}(\{\Omega_{\g_n}\}_{n=1}^{N_\tau},\Omega_{\bm\gamma},\Omega_{\bm\rho},\Omega_{\alpha},\Omega_{\z})$ has a lower bound, the sequence generated by the above iterations converges to a limit. Together with the fact that (\ref{parameter1})-(\ref{parameter7}) have unique solutions, we are able to establish that the above limit is a stationary point by using the Theorem~2-b in \cite{Razaviyayn2014}.
\section{}
\begin{figure}
\center
\begin{tikzpicture}[scale=0.78]
\begin{semilogyaxis}[xlabel={Number of iterations},font=\large,
ylabel={NMSE},
title={\large(a) OGVBI},
grid=major,
legend style={at={(0.782,1)},
anchor=north,legend columns=1},xmin=1,xmax=100,ymin=0.001,ymax=1]

%\addplot[color=blue, line width=1pt] file {Runtime/OG_-10dB.txt};
%\addlegendentry{\scriptsize \,SNR = -10 dB}
\addplot[color=red, line width=1pt] file {Runtime/OG_0dB.txt};
\addlegendentry{\small SNR = 0 dB}
\addplot[color=green, line width=1pt] file {Runtime/OG_10dB.txt};
\addlegendentry{\small \textcolor[rgb]{1.00,1.00,1.00}{1}SNR = 10 dB}
\addplot[color=black, line width=1pt] file {Runtime/OG_20dB.txt};
\addlegendentry{\small \textcolor[rgb]{1.00,1.00,1.00}{1}SNR = 20 dB}
\end{semilogyaxis}
\end{tikzpicture}
\begin{tikzpicture}[scale=0.78]
\begin{semilogyaxis}[xlabel={Number of iterations},font=\large,
ylabel={NMSE},
title={\large(b) Vector-OGVBI},
grid=major,
legend style={at={(0.782,1)},
anchor=north,legend columns=1},xmin=1,xmax=100,ymin=0.001,ymax=1]
%\addplot[color=blue, line width=1pt] file {Runtime/SBL_-10dB.txt};
%\addlegendentry{\scriptsize \,SNR = -10 dB}
\addplot[color=red, line width=1pt] file {Runtime/SBL_0dB.txt};
\addlegendentry{\small SNR = 0 dB}
\addplot[color=green, line width=1pt] file {Runtime/SBL_10dB.txt};
\addlegendentry{\small \textcolor[rgb]{1.00,1.00,1.00}{1}SNR = 10 dB}
\addplot[color=black, line width=1pt] file {Runtime/SBL_20dB.txt};
\addlegendentry{\small \textcolor[rgb]{1.00,1.00,1.00}{1}SNR = 20 dB}
\end{semilogyaxis}
\end{tikzpicture}
\begin{tikzpicture}[scale=0.78]
\begin{semilogyaxis}[xlabel={Number of iterations},font=\large,
ylabel={NMSE},
title={\large(c) Fast-VBI},
grid=major,
legend style={at={(0.782,1)},
anchor=north,legend columns=1},xmin=1,xmax=100,ymin=0.001,ymax=1]
%\addplot[color=blue, line width=1pt] file {Runtime/Fast_-10dB.txt};
%\addlegendentry{\scriptsize \,SNR = -10 dB}
\addplot[color=red, line width=1pt] file {Runtime/Fast_0dB.txt};
\addlegendentry{\small SNR = 0 dB}
\addplot[color=green, line width=1pt] file {Runtime/Fast_10dB.txt};
\addlegendentry{\small \textcolor[rgb]{1.00,1.00,1.00}{1}SNR = 10 dB}
\addplot[color=black, line width=1pt] file {Runtime/Fast_20dB.txt};
\addlegendentry{\small \textcolor[rgb]{1.00,1.00,1.00}{1}SNR = 20 dB}
\end{semilogyaxis}
\end{tikzpicture}
\begin{tikzpicture}[scale=0.78]
\begin{semilogyaxis}[xlabel={Number of iterations},font=\large,
ylabel={NMSE},
title={\large(d) Proposed},
grid=major,
legend style={at={(0.782,1)},
anchor=north,legend columns=1},xmin=1,xmax=100,ymin=0.001,ymax=1]
%\addplot[color=blue, line width=1pt] file {Runtime/Our_-10dB.txt};
%\addlegendentry{\scriptsize \,SNR = -10 dB}
\addplot[color=red, line width=1pt] file {Runtime/Our_0dB.txt};
\addlegendentry{\small SNR = 0 dB}
\addplot[color=green, line width=1pt] file {Runtime/Our_10dB.txt};
\addlegendentry{\small \textcolor[rgb]{1.00,1.00,1.00}{1}SNR = 10 dB}
\addplot[color=black, line width=1pt] file {Runtime/Our_20dB.txt};
\addlegendentry{\small \textcolor[rgb]{1.00,1.00,1.00}{1}SNR = 20 dB}
\end{semilogyaxis}
\end{tikzpicture}
\caption{NMSEs of various methods versus the number of iterations with $L=40$ and $N_{BS}=40$.}
\end{figure}
This appendix provides a computational complexity comparison between the proposed method and various baselines. First, we outline the proposed method in Algorithm~1 and give its detailed computational complexity analysis as follows:
 \begin{itemize}
   \item In Step 3-a), the computational complexities in calculating $\bm\mu_n$ and $\bm\Sigma_n$ are $\mathcal{O}(M_\theta^2)$ and $\mathcal{O}(N_{BS}^2 M_\theta)$ for each delay tap $n$, respectively. Therefore, the computational complexities in calculating $\{\bm\mu_n\}_{n=1}^{N_\tau}$ and $\{\bm\Sigma_n\}_{n=1}^{N_\tau}$, summed over all $N_\tau$ delay taps, amount to $\mathcal{O}(N_\tau M_\theta^2)$ and $\mathcal{O}(N_\tau N_{BS}^2 M_\theta)$ per iteration, respectively.
   \item In Step 3-b), the computational complexities in calculating $\{c_{\gamma_m},d_{\gamma_m}\}_{m=1}^{M_\theta}$, $\{c_{\rho_n},d_{\rho_n}\}_{n=1}^{N_\tau}$ and $\{\hat z_{m,u}\}_{m=1,u=-1}^{M_\theta,1}$ are negligible (as no matrix operations are involved), and the computational complexity in calculating $\{c_{\alpha},d_{\alpha}\}$ is $\mathcal{O}(N_\tau N_{BS} M_\theta^2)$ per iteration.
   \item In Step 3-c), the computational complexities in calculating $\bm\theta$ and $\bm\kappa$ are $\mathcal{O}(N_\tau N_{BS} M_\theta^2)$ and $\mathcal{O}(N_\tau L^3)$ per iteration, respectively.
 \end{itemize}
Therefore, the total computational complexity of the proposed method is $\mathcal{O}(IN_\tau N_{BS}^2 M_\theta)$, where $I$ represents the number of iterations.

Table~I provides the comparison of computational complexities of various methods.
It is evident that: (i) the proposed method and Fast-VBI have identical orders of computational complexity, since the computational complexities of updating $q(\bm\gamma)$, $q(\bm\rho)$ and $q(\z)$ are negligible in Step 3-b); (ii) in contrast to Vector-OGVBI which utilizes the standard VBI factorization, the proposed method achieves a significant reduction in computational complexity by adopting the independent VBI factorization; (iii) although the proposed method and Fast-VBI display comparable computational complexities to $\ell_1$-norm and LS-based methods, the latter two methods cannot exploit channel sparsity to enhance their performance; and (iv) although OGVBI has the lowest computational complexity, it fails to leverage the sparsity exhibited in the delay domain, resulting in a substantial performance loss.
\begin{algorithm}[htbp]
\caption{{Independent VBI Factorization-Based OTFS Channel Estimation Algorithm}}
\begin{enumerate}
  \item Input: $\Y$ and $\x$.
  \item Initialization: $\hat\gamma_m=1$, $\forall m$; $\hat\rho_n=1$, $\forall n$; $\hat\alpha=1$; $\hat z_{m,u}=\frac{1}{3}$, $\forall m,u$; $c=d=0.001$.
  \item Repeat the following until convergence:
  \begin{itemize}
    \item[a)] Calculate $\{\bm\mu_n\}_{n=1}^{N_\tau}$ and $\{\bm\Sigma_n\}_{n=1}^{N_\tau}$ to update $\{q(\g_n)\}_{n=1}^{N_\tau}$ in (\ref{update1}).
    \item[b)] Calculate $\{c_{\gamma_m},d_{\gamma_m}\}_{m=1}^{M_\theta}$, $\{c_{\rho_n},d_{\rho_n}\}_{n=1}^{N_\tau}$, $\{c_{\alpha},d_{\alpha}\}$, and $\{\hat z_{m,u}\}_{m=1,u=-1}^{M_\theta,1}$ to update $q(\bm\gamma)$, $q(\bm\rho)$, $q(\alpha)$, and $q(\z)$ in (\ref{update2})-(\ref{update5}), respectively.
    \item[c)] Update $\bm\theta$ and $\bm\kappa$ using (\ref{grid-refine}) and (\ref{Doppler-refine}), respectively.
  \end{itemize}
  \item Output: $\{\bm\mu_n\}_{n=1}^{N_\tau}$, $\bm\theta$ and $\bm\kappa$.
\end{enumerate}
\end{algorithm}
\begin{table}[htbp]
\centering
\caption{\scshape{Computational Complexities of Various Methods}}
\begingroup
\begin{tabular}{|c|c|c|}

   \hline
  Method & Computational complexity  \\

   \hline
   LS &$\mathcal{O}(LN_\tau^2 N_{BS}  M_\theta^2 )$ \\

  \hline

   $\ell_1$-norm &$\mathcal{O}(LN_\tau^2 N_{BS}^2  M_\theta )$ \\

  \hline

   OGVBI &$\mathcal{O}(IN_{BS}^2 M_\theta)$\\
  \hline
   Vector-OGVBI &$\mathcal{O}(IL^2 N_{BS}^2 N_{\tau} M_{\theta})$\\
  \hline
   Fast-VBI &$\mathcal{O}(IN_\tau N_{BS}^2 M_\theta)$\\
  \hline
Proposed &$\mathcal{O}(IN_\tau N_{BS}^2 M_\theta)$\\
   \hline
\end{tabular}
\endgroup
\end{table}

\section{}

This appendix compares the maximum number of iterations for convergence across different schemes to enable a straightforward comparison of convergence speed and overall computation time with the baseline schemes.
Fig.~6 shows the NMSEs of various methods versus the number of iterations, where the
pilot length is set to $L=40$ and the antenna number is set to $N_{BS}=40$. It is observed that the compared methods (OGVBI,  Vector-OGVBI and Fast-VBI) typically converge within 50 iterations, whereas the proposed method converges within 60 iterations. The slower convergence speed of the proposed method is due to its ability to achieve a lower NMSE than the other methods, which naturally requires more iterations. Additionally, it is evident that even when the proposed method has not yet converged (e.g., at the 50th iteration), it still achieves better NMSE performance than the fully converged compared methods.

\bibliographystyle{IEEEtran}
\bibliography{reference}

\end{document}